\documentclass[twocolumn,showpacs,amsmath,amssymb]{revtex4}
\usepackage{graphicx}
\usepackage{dcolumn}
\usepackage{bm}
\usepackage[dvips]{epsfig}
\newcommand{\NC}[3]{ #2 {\em Nuovo Cimento\ }{\bf #1} #3}

\begin{document} 

\newcommand{\be}{\begin{equation}}
\newcommand{\ee}{\end{equation}}
\newcommand{\ba}{\begin{eqnarray}}
\newcommand{\ea}{\end{eqnarray}}
\newcommand{\nn}{\nonumber\\}
\newcommand{\n}[1]{\label{#1}}
\newcommand{\hb}{\widehat{\Box}}

\newcommand{\bn}{{\mbox{\boldmath $\nabla $}}}
\newcommand{\BM}[1]{{\mbox{\boldmath $#1$}}}
\newcommand{\BF}[1]{\mbox{\bf #1}}
\newcommand{\BFE}[1]{\mbox{\bf e}_{\hat{#1}}}
\newcommand{\hn}{\hat{\nabla}}

\newcommand{\eq}[1]{(\ref{#1})}
\newcommand{\ind}[1]{\mbox{\tiny #1}}

\newcommand{\hh}{\, ,\hspace{0.5cm}}
\newcommand{\hhh}{\, ,\hspace{0.2cm}}

\newpage

\title{Black Holes in Higher Dimensional Gravity Theory
with Quadratic in Curvature Corrections}

\author{Valeri P. Frolov$^*$  and Ilya L. Shapiro$^\dagger$}

\affiliation{
  \medskip
  $^*$Theoretical Physics Institute, 
  Department of Physics, University of Alberta\\
  Edmonton, AB, Canada, T6G 2J1\\
  {\rm E-mail: \texttt{frolov@phys.ualberta.ca}}
  \medskip
}

\affiliation{
  \medskip
  $^\dagger$ Departamento de F\'{\i}sica, ICE,
Universidade Federal de Juiz de Fora \\
CEP: 36036-330, Juiz de Fora, MG, Brazil\\
 {\rm E-mail: \texttt{shapiro@fisica.ufjf.br}}}

\date{\today}

\begin{abstract}
Static spherically symmetric black holes are discussed
in the framework of higher dimensional gravity with quadratic 
in curvature terms. Such terms naturally arise as a result of 
quantum corrections induced by quantum fields propagating in 
the gravitational background. We focus our attention on the 
correction of the form
${\cal C}^2=C_{\alpha\beta\gamma\delta} C^{\alpha\beta\gamma\delta}$.
The Gauss-Bonnet equation in four-dimensional ($4D$) spacetime 
enables one to reduce this term in the action to the terms quadratic 
in the Ricci tensor and scalar curvature. As a result the Schwarzschild
solution which is Ricci flat will be also a solution of the theory
with the Weyl scalar ${\cal C}^2$ correction. An important new feature
of the spaces with dimension $D > 4$ is that in the presence of the
Weyl curvature-squared term a solution necessary differs from the
corresponding `classical' vacuum Tangherlini metric. This difference 
is related to the presence of {\em secondary} or {\em induced} hair. 
We explore how the Tangherlini solution is modified by `quantum 
corrections', assuming that the gravitational radius $r_0$ is much 
larger than the scale of the quantum corrections. We also demonstrated 
that finding a general solution beyond the perturbation method can be 
reduced to solving a single third order ODE (master equation). 
\end{abstract}

\pacs{04.50.-h, 
04.60.Bc, 
04.70.-s, 
97.60.Lf  
\hfill   Alberta.-Thy-08-09
}

\maketitle

\section{Introduction}\label{s1}

The Einstein-Hilbert action is the simplest possible gravitational 
action. It can be easily generalized to the higher dimensional 
models of gravity
\be
W = \frac{1}{16\pi} \int d^D x \sqrt{-g}\, R\, .
\ee
Black hole solutions with the spherical topology of the horizon in 
this theory are well known now. The most
general solution (in the presence of a cosmological constant) is a
so-called Kerr-NUT-(A)dS spacetime \cite{Pope}. In the absence of
rotation and for vanishing $NUT$ parameters and the cosmological
constant, this solution reduces to the Tangherlini metrics \cite{Tang}
\be
ds^2 = - A dt^2 + \frac{dr^2}{A} + r^2 d\omega_n^2\hh
A=1 - \Big(\frac{r_0}{r}\Big)^{n-1}\, .
\ee
For a $D$-dimensional spacetime, $n=D-2$, and $d\omega_n^2$ is the
metric of a unit round sphere $S^n$.

Modifications of the Einstein-Hilbert action are commonly 
considered in the modern literature.  Besides the needs of a 
purely phenomenological description of gravity in different 
models, there exists a more fundamental reason for considering 
a generalized Einstein gravity. It is well known that quantum 
corrections in a gravitational field can be described by using
the DeWitt's effective action formalism \cite{DW}. In a general 
case, such an effective action contains higher in curvature 
corrections to the Einstein-Hilbert Lagrangian, as well as 
non-local contributions (see, e.g., review 
\cite{PoImpo} and references therein).
Higher in curvature corrections were used for example in the
Starobinsky model of inflation \cite{star}. The higher curvature
terms naturally arise in the string theory (see e.g. \cite{Pol}
and references therein). One of the most interesting aspects
connected to the introduction of the higher in curvature
effective actions is a possibility to resolve a curvature
singularity problem. Such singularities, according to the Penrose
and Hawking \cite{Pen,Hawk,PH} theorems always exist in cosmology
(initial singularity) and inside black holes (final singularity).
It is well known that the interior of (non-rotating) black
hole in the vicinity of the singularity is similar to a
contracting anisotropic universe and it exhibits the Kasner type
behavior with infinitely growing curvature invariants. In the
models where the curvature is limited by some (say Planckian)
values, one may expect newly born universes instead of the
singularity formation (see, e.g. \cite{FMM,BF}). For more
general discussion of the quantum effects in black holes see e.g.
the book \cite{FN} and references therein.

Let us notice that the consistent consideration of quantum
effects in gravity requires an introduction of the higher derivative
terms, even at the semiclassical level (see
\cite{birdav,book} for the introduction and \cite{PoImpo}
for the recent review of semiclassical approach).
One of the first papers where the quantum curvature
corrections were considered in the gravitational collapse model was
\cite{FV}. In particular, it was shown that a collapse of a null shell
with the mass smaller that the Planckian mass does not create a black
hole. In other words, in this problem there exists a mass gap for the
black hole formation. In the modern language it means that the black
hole formation in the gravitational collapse of null shells is a
first order phase transition. Later, black holes in the theories with
higher in curvature terms were considered, e. g. in \cite{HigherBH}.

In the four-dimensional ($4D$) case the lowest in curvature
quadratic corrections to the Einstein gravity do not modify
the Schwarzschild solution. The reason is the following.
In $4D$ case there exists the Gauss-Bonnet relation
\be
C_{abcd}C^{abcd} - 2\Big(R_{ab}R^{ab} - \frac13\,R^2\Big)
= \mbox{topological term}\, .
\ee
As a consequence, the general $4D$ action with quadratic in
curvature corrections can be always written as
\be
\n{Rab}
W = \frac{1}{16\pi}\int d^4 x
\sqrt{-g}\,\left( R +a R_{ab}R^{ab}+b R^2\right)\, .
\ee
It is clear that any vacuum solution of the Einstein equations,
$R_{ab}=0$ is, at the same time, a solution of the theory \eq{Rab}.
For this reason, in order to study the quantum gravity corrections
we need to use higher in curvature Lagrangians. One of the
interesting option is to consider $4D$ theories with the
corrections of the form
$f({\cal C}^2)$, which in many respects are similar to the $f(R)$
theories in the cosmological models.

In this paper we use another approach. We consider higher
dimensional spherically symmetric black holes with quadratic
in the curvature corrections. Since the Gauss-Bonnet argument
does nor work directly in the higher dimensions, the quadratic
in the Weyl tensor corrections necessarily modify the background
Tangherlini solution. This results in the creation of what is
called {\em secondary} or {\em induced hair} \cite{CPW1,CPW2}.
In the present paper we focus our attention on the study of
these `hair'. Namely, we consider static spherically symmetric
vacuum solutions of the gravitational theory with ${\cal C}^2$
corrections. Using the reduced action approach we derive
equations for such spacetimes (sections~II and III). In
section~IV we demonstrate that in the special (seminull)
coordinated these equations can be reduced to a single third
order ordinary differential equation (ODE) for a quantity,
connected to the Weyl curvature. Linearized version of the
field equations and their static solutions are discussed in
section~V. In order to obtain corrections to the Tangherlini
metric describing the induced hair generated by the ${\cal C}^2$
correction, we use the iteration procedure developed in section~VI.
This section also contains the numerical results. Specific
properties of the $4D$ case are discussed in section~VII.
Section~VIII contains discussion and lists some of the
perspective problems.

Throughout the paper we use the sign conventions adopted in
\cite{MTW}.

\section{Reduced action}
\label{s2}

In a $D$-dimensional spacetime with metric
\be\n{1}
ds^2=g_{\mu\nu}dx^{\mu}dx^{\nu}
\ee
we consider a gravitational action of the form
\ba\n{2}
W &=& {1\over 16\pi}\int d^D x \sqrt{-g}\,
\Big(R\,-\,{a\over 2}{\cal C}^2\Big)\,,
\\
\mbox{where} &\,&
{\cal C}^2
= C_{\alpha\beta\gamma\delta}C^{\alpha\beta\gamma\delta}\,.
\ea
Here $R$ and $C_{\alpha\beta\gamma\delta}$ are the Ricci scalar
and the Weyl tensor, respectively.
The constant $a$ has the dimensionality of  $\mbox{[length]}^2$,
and, for physically realistic case, it
is positive. The reason is that this  sign corresponds to the
positively defined theory, e.g.,  to the positive energy of
the massless mode (graviton) in the flat space limit. This
requirement is also relevant because it provides the correct
low-energy limit of the theory \cite{Stelle,book}. The constant
$a$ can be interpreted as a square of the characteristic fundamental
scale $l$ for an appropriate quantum gravity theory. In a `standard
quantum gravity' $l$ is of the order of the Planckian length. In the
popular recently models with large extra dimensions the corresponding
to $l$ fundamental energy scale is of order of TeV. We shall study
spherically symmetric black hole solutions. The characteristic scale
of such a solution is its gravitational radius $r_0$. This
dimensional parameter can be chosen as a general metric scale.
After this in the theory \eq{2} there remains only one essential
dimensionless parameter $l/r_0$ (or, equivalently, $a/r_0^2$).
One can expect that when $r_0\gg l$, the Weyl scalar
(Weyl curvature-squared) term plays
the role of a small correction. We shall focus on this case later 
on, but at the initial part of the work there is no need to make
additional assumptions about this parameter.

Our purpose is to explore static spherically symmetric
solutions of the theory formulated above.
The corresponding metric can be written in the form
\be
ds^2=r^2 d\bar{s}^2=r^2\, (d\gamma^2+d\omega_n^2)\, ,
\ee
where $n=D-2$ and $d\omega_n^2$ is the metric on a unit sphere $S^n$
\be
d\omega_n^2=\omega_{ij}d\theta^i d\theta^j\hh
i,j,\ldots=2,\ldots,n+1\, .
\ee
Quantity $r$ is a scalar function on a $2D$ manifold with the metric
\be
d\gamma^2=\gamma_{ab} dx^a dx^b\hh a,b,\ldots=0,1\, .
\ee
For this class of the metrics the action \eq{2} can be reduced to
the $2D$ action. Because of the spherical symmetry neither $R$
nor ${\cal C}^2$ depend on the angle variables. Using the relation
\be
\sqrt{-g}=r^{n+2} \sqrt{-\gamma}\sqrt{\omega}\, ,
\ee
and integrating over the unit sphere one obtains
\ba\n{}
W &=& {{\cal A}_n\over 16\pi}{\cal W}\hh
{\cal W}=\int d^2 x \sqrt{-\gamma} {\cal L}\, ,\\
{\cal L}&=&r^{n+2}\, \Big(R-{a\over 2}{\cal C}^2\Big)\,.
\ea
Here
$$
{\cal A}_n = \frac{2 \pi^{(n+1)/2}}{\Gamma \big(\frac{n+1}{2}\big)}
$$
is a surface area of the $n$-dimensional unit sphere $S^n$.

We denote by symbols with bars the objects constructed for the
metric $d\bar{s}^2$. Since this metric is a direct sum of the
two metrics, $d\gamma^2$ and $d\omega^2$, the curvature for
it is also a direct sum of two curvatures and it has the
following non-vanishing components
\ba\n{bar}
\bar{R}_{abcd} &=&
{1\over 2}\,\hat{R}\,(\gamma_{ac}\gamma_{bd}-\gamma_{ad}\gamma_{bc})
\,,
\nonumber
\\
\bar{R}_{ijkl}&=&\omega_{ik}\omega_{jl}-\omega_{il}\omega_{jk}\,.
\ea
We also have
\ba
\bar{R}_{ac}&=&{1\over 2}\hat{R}\gamma_{ac}\n{bar1}\hh
\bar{R}_{ik}=(n-1)\omega_{ik}\, \n{bar2}
\nonumber
\\
\bar{R}&=&\hat{R}+n(n-1)\, .
\n{bar3}
\ea

We denote by symbols with hats the objects constructed for the
$2D$ metric $d\gamma^2$. In particular, $\hat{R}$ and
$\hat{{\cal C}}^2$
denote the Ricci scalar and the Weyl-square invariant, calculated
for the metric $\gamma_{ab}$. It is easy to show that
(see, e.g., \cite{Stud})
\ba
R&=&r^{-2}\, \left[ \hat{R}+n(n+1)-2(n+1){\hb r\over r}\right.
\nonumber
\\
&-&\left. (n+1)(n-2){(\hat{\nabla}r)^2\over r^2} \right]\, .
\ea

The conformal invariance of the Weyl tensor implies
\be
{\cal C}^2=r^{-4} \bar{{\cal C}}^2\, .
\ee
To calculate $\bar{{\cal C}}^2$ we use a relation
\be
\n{c2}
\bar{{\cal C}}^2=
\bar{R}_{\alpha\beta\gamma\delta}\bar{R}^{\alpha\beta\gamma\delta}
-{4\over n}\bar{R}_{\alpha\beta}\bar{R}^{\alpha\beta}
+{2\over n(n+1)}\bar{R}^2\, .
\ee

By using \eq{bar}--\eq{bar3} one can obtain
\ba
\bar{R}_{\alpha\beta\gamma\delta}^2 &=&
{\hat R}^2 + 2n(n-1)\, ,\\
\bar{R}_{\alpha\beta}^2 &=&
\frac12\,{\hat R}^2 + n(n-1)^2\, ,\\
\bar{R}^2 &=&\Big[ {\hat R} + n(n-1)\Big]^2\, .
\ea
Then we find
\be
\bar{{\cal C}}^2={n-1\over n+1}(\hat{R}+2)^2\, .
\nonumber
\ee

Thus the dimensionally reduced action on the $2D$
manifold with the metric $\gamma_{ab}$ is
\ba
\n{ra}
{\cal W} = {\cal W}_{(c)} + {\cal W}_{(q)}
= \int d^2 x \sqrt{-\gamma}\,
\big[{\cal L}_{(c)}+{\cal L}_{(q)}\big],
\,\,
\\
{\cal L}_{(c)} =
r^n \hat{R}+n(n-1) r^n + n(n+1)r^{n-2}(\hat{\nabla
r})^2 ,
\\
{\cal L}_{(q)} =  - \frac{a(n-1)\,r^{n-2}}{2(n+1)}\,(\hat{R}+2)^2 \,.
\qquad\qquad\qquad\,\,
\ea

\section{Field equations}

The reduced action ${\cal W}$ is a functional of the $2D$
metric $\gamma_{ab}$ and of a scalar function $r$. Its variation
has the form
\be\n{var}
\delta {\cal W} = \int d^2 x \sqrt{-\gamma}\,
\left[ {\cal P}\delta r +{\cal G}^{ab}\delta\gamma_{ab}\right]\, .
\ee
Consider a coordinate transformation generated by a vector field
$\xi^a$
\be
x^a\to x^a +\xi^a\, .
\ee
Under this transformation $r$ and $\gamma_{ab}$ transform as follows
\be\n{delta}
\delta_{\xi}r=-\xi^a r_{,a}\hh
\delta_{\xi} \gamma_{ab}=-2\hat{\nabla}_{(a}\xi_{b)}\, .
\ee
Since ${\cal W}$ is invariant under this transformation one has
$\delta_{\xi}{\cal W}=0$. Using \eq{var} and \eq{delta} one obtains
the following {\em Noether identity}
\be\n{ni}
2 \hn_b{\cal G}^{b}_{\ a}-{\cal P}\hat{\nabla}_a r=0\, .
\ee

To obtain the field equations for the reduced action we use
the following relations
\ba
&&\delta\sqrt{-\gamma}={1\over 2}\sqrt{-\gamma} \gamma^{ab}
\delta\gamma_{ab}\, ,
\nonumber
\\
&&
\int d^2 \sqrt{-\gamma} (\delta \hat{R}) S
= \int d^2 \sqrt{-\gamma}\,\delta\gamma_{ab}\,
\left[\hat{\nabla}^a \hat{\nabla}^b S \right.
\nonumber
\\
&& \quad\quad\quad\quad\left.-\gamma^{ab} \hb S
- \hat{R}^{ab} S\right]\,.
\ea

The variation of \eq{ra} with respect to $r$ gives
\ba\n{eqr1}
{\cal P}&=&{\cal P}_{(c)}+{\cal P}_{(q)}\, ,\\
{\cal P}_{(c)}&=& n r K
-n(n+1)r^{n-3}[(n-2)(\hat{\nabla}r)^2\nonumber\\
&+&2 r\hb r -(n-2)r^2]\,
,\n{eqr2} \\
{\cal P}_{(q)}&=&-{a(n-1)(n-2)\over 2(n+1)r^{n-1}}K^2\, .\n{eqr3}
\ea
Here and later we use subscripts $c$ and $q$ for quantities
connected with  `classical' and `quantum' parts of the action,
respectively. Similarly, the variation of \eq{ra} with respect
to $\gamma_{ab}$ gives
\ba
\n{eqg1}
{\cal G}^{ab}
= {\cal G}^{ab}_{(c)}+{\cal G}^{ab}_{(q)}\,,
\ea 
where
\ba
{\cal G}^{ab}_{(c)}
&=& nr^{n-2} \Big\{ r \hn^{a}\hn^{b}r-2\hn^ar\hn^b r
\n{eqg2}
\\
&+& {1\over 2}\gamma^{ab} \big[ (n-1)r^2 -(n-3)(\hn r)^2-2r\hb r 
\big] \Big\}
\nonumber
\ea
and  
\ba
{\cal G}^{ab}_{(q)} &=& - {a(n-1)\over (n+1)}
\Big[ \hn^a\hn^b K + \gamma^{ab} (K -\hb K
\nonumber
\\
&-& {1\over 4} r^{2-n}K^2)\Big]
\,.
\n{eqg3}
\ea
In the above relations  we use, instead of $\hat{R}$, the
related quantity,
\be
K = r^{n-2}(\hat{R}+2)\,,
\ee
which is directly connected to the Weyl scalar
in the `physical' spacetime
\be
|K|=\sqrt{n+1\over n-1}r^n ({\cal C}^2)^{1/2}\, .
\ee

Furthermore, it is easy to show that
\be
\hn_b {\cal G}^{b}_{(q) a}=-{a(n-2)(n-1)\over 4 (n+1)r^{n-1}}
K^2 \hn_a r\, .
\ee
This relation allows one to check the validity of the Noether
identity \eq{ni} for the `quantum' part of the action. This
serves as a good test of the rather long calculations required
to derive relations \eq{eqr1}-\eq{eqg3}.

To summarize, the vacuum field equations for the reduced action
\eq{ra} are
\be
{\cal P}=0\hh {\cal G}_{ab}=0\, .
\ee

%
\section{Explicit form of the field equations}

In the absence of the Weyl term in the action one can show 
that any spherically symmetric vacuum solution of the higher 
dimensional Einstein equations is static (a {\em generalized 
Birkhoff's theorem}). In the presence of the Weyl term this 
is not true anymore. In the present work we restrict ourselves 
by considering static solutions of the theory \eq{1}, that 
is a special subclass of its solutions.

Let us describe the form of the action which is well adapted 
to our problem. We use the coordinate freedom $r\to \tilde{r}(r)$ 
in order to put $g_{rr}=0$, and write
 the  metric $\gamma_{ab}$ in the form
\be
d\gamma^2 = -{A \over r^2}\,e^{2C} dv^2
+ {2\over r^2}\,e^{C} dv dr\, ,
\label{metric}
\ee
where $A=A(r)$ and $C=C(r)$ are some unknown functions.
The corresponding `physical' metric is
\be
\n{seminull}
ds^2=-Ae^{2C} dv^2 +2 e^{C} dv dr+r^2 d\omega^2_n\, .
\ee
We assume that the spacetime is asymptotically flat. This implies 
that $A(r=\infty)=1$. The function $C(r)$ takes a finite value 
$\tilde{C}$ at infinity. This value depends on the normalization
of the advanced time coordinate $v$. We choose its normalization 
so that $\tilde{C}=0$. For this choice $\xi=\partial_v$, which 
is a Killing vector, has a canonical normalization at the 
infinity: $\xi^2=-1$.

The event horizon, if it exists, is located at the surface where 
$A=0$. The metric \eq{seminull} in the coordinates
$(v,r)$ is regular at this surface, provided $C$ is regular there. In particular, this property  enables one
easily to consider modifications of the Schwarzschild black hole
solution due to the quantum corrections, in a case when the
latter are small and can be treated as perturbations
(see section~4). In the presence of a black
hole, its event horizon is located at the point $r_H$ where
the function $A$ vanishes. The surface gravity $\kappa$ of
a static black hole is defined as
\be
\kappa=\Big({1\over 2}\xi_{a;b}\xi^{a;b}\Big)^{1/2}\, ,
\ee
where $\xi=\partial_v$ is the Killing vector. For the
metric \eq{seminull} this expression can be rewritten in a more
simple form,
\be
\kappa={1\over 2} (e^{C}A')\big|_{r=r_H}\, .
\n{kappa}
\ee
The Hawking temperature of such a black hole is \
$T_H=2\pi/\kappa$. For the unperturbed metric, when $a=0$, one has
\be
\kappa_0={n-1\over 2 r_0}\, .
\ee

Calculations using GRTensor program allows one to obtain the
following relations. The non-vanishing components of the
Christoffel symbols are
\ba
{\Gamma _{v\,v}}^{v} &=& - {\Gamma _{r\,v}}^{r}=V
\hh
{\Gamma _{v\,v}}^{r} = AVe^{C}\,,
\nonumber \\
{\Gamma _{r\,r}}^{r} &=& C'-{2\over r}\,,
\ea 
where
\ba
V={1\over 2r}(A'r+2AC'r-2A)\,e^{C}\, .
\ea
One also has ${\cal G}_{vv}={\cal G}_{vr}$, due to algebraic
properties of the metric (\ref{metric}) and corresponding
components of the Ricci tensor. Finally, it proves useful to
introduce a parameter $p$, related to $a$ as
$$
p = - \frac{a}{2(n+1)}\,.
$$
Notice that according to our assumption this parameter is 
{\em negative}. The non-vanishing field equations can be 
presented in the form
\ba
&&
(n-1)p\Big[4r^2AK''+ 2r(rA'+2A)K'-4K+{K^2\over r^{n-2}}\Big]
\nonumber
\\
&&
+ nr^n[rA'+(n-1)(A-1)]\,=\,0\, ,
\n{vr}
\\
\nonumber\\
&&
nr^n C'\,-\,2(n-1)p\big[rK''-rC'K'+2K'\big]=0\, ,\n{rr}
\\
\nonumber\\
&&
(n-1)(n-2)\,pK^2 r^{2(1-n)}
\,-\,n\,\big[(n-2)(n+1)(A-1)\nonumber\\
&& + 2(n+1)r(A C'+A') - K\,r^{2-n}\big]=0\,,
\n{P}\\
\nonumber\\
&&
K = - r^{n-2}\big[
3r^2A'C'-2-2r(A'+AC')\nonumber\\
&&+2A + r^2A'' + 2Ar^2(C''+{C'}^2)\big]\, .\n{k}
\ea
The equations \eq{vr} and \eq{rr} follow from $(vr)$ and
$(rr)$ components of the equation ${\cal G}_{ab}=0$, while
\eq{P} follows from the equation ${\cal P}=0$. The last
equation is the definition of $K$.

A remarkable property of the system of equations \eq{vr}--\eq{P} 
is that they can be reduced to a single third order ODE.
We notice first that this system does not contain the function 
$C(r)$, but only its derivatives. The \eq{rr} enables one to 
express $C'$ as a function of $K$ and its derivatives
\be
C'\,=\,\frac{2(n-1)\,p\,(rK''+2K')}{nr^n+2(n-1)prK'}\, .
\ee
After substituting this expression into \eq{vr} and \eq{P},
one obtains two linear equations for $A$ and $A'$. Solving
these equation one determines $A$ and $A'$ as functions
of $K$ and its derivatives up to the second order.
One can present the expression for $A$ in the form
\be\n{AAA}
A={S_1K'+S_0\over n(T_2K''+T_1K'+T_0)}\, ,
\ee
where
\ba
S_1 &=& -2p(n-1) \times
\nonumber
\\
&\times&
\left[ n(n-2)(n+1)r -p(n^3-3n+2){K^2\over
r^{2n-3}}\right]\, ,
\nonumber\\
S_0 &=&
n\left[ 8(n^2-1)pK - n r^2K - \frac{p(n-1)(n+4)K^2}{r^{n-2}}\right],
\nonumber\\
T_2&=&4(n^2-1)pr^2 \hh \quad
T_1=-2p(n^2-1)(n-2)r\, ,
\nonumber\\
T_0&=&n^2(n+1)r^n\,.
\nonumber
\ea
A similar expression is valid for $A'$.
Taking a derivative of $A$ given by \eq{AAA} and putting 
it equal to $A'$, one
arrives at a single ordinary differential equation of the
third order for $K$. We call it {\em master equation}.
We have obtained this equation using the MAPLE program
but do not reproduce it here since it is too bulky.

\section{Linearized equations and their solutions}

Since the equations of our interest are rather complicated,
it is worthwhile to consider first their linearized form. This
exercise is especially useful, because it helps to control
the correct classical ($p\to 0$) limit in the region
far from the center $r=0$. Let us denote
\be
A=1+\alpha(r)\hh
C=\beta(r)\hh
K=\nu(r)\, .
\ee
We regard the functions $\alpha$, $\beta$, and $\nu$ as
small perturbations and hence, in the course of our
calculations, we shall keep only those terms
which are linear in these functions.
The linearization of equations
\eq{vr}--\eq{P} gives
\ba
&&
n(n-1)\alpha+nr\alpha'
\\
\nonumber
&& - 4(n-1)p \,r^{-n}\big(\nu-r\nu'-r^2\nu''\big)=0\, ,
\n{eq1}
\\
\nonumber
\\
&& 2(n-1)p(r\nu''+2\nu')-nr^n\beta'=0\, ,\n{eq2}
\\
\nonumber
\\
&& \big[(n-2)\alpha+2r(\alpha'+\beta')\big]
- \frac{\nu r^{2-n}}{n+1}=0 ,
\n{eq3}
\ea 
\ba 
\nu r^{2-n}
+ r^2(\alpha''+2\beta'')-2r(\alpha'+\beta')+2\alpha=0 .
\n{eq4}
\ea
It is easy to check that the last equation \eq{eq4} follows 
from the first 3 equations. This system of equations can 
be reduced to the single third order ODE (master equation) 
for the function $\nu$.
This can be done by adopting the procedure described in the 
previous section to linearized case. Namely, let us find
$\beta'$ from \eq{eq2} and substitute it into \eq{eq1} and
\eq{eq3}. Use these two equations to find $\alpha$ and $\alpha'$.
Substitute the obtained expression for $\alpha'$ into \eq{1}
and differentiate the obtained equation with respect to $r$.
After this, again substitute $\alpha'$ in the equation.
These operations produce the following {\em master} equation
which contains only $\nu$ and its derivatives~\footnote{The
same equation can be obtained by linearizing the general 
master equation which has been mentioned by the end of the 
previous section.}
\ba
&& nr^2(\nu+r\nu')+4(n^2-1)(n+2)p(\nu-r\nu')
\nonumber\\
&& +4p(n^2-1)r^2[r\nu'''-(n-1)\nu'']=0\, .
\n{master nu}
\ea
Before we solve this equation, it proves useful to perform
a linear change of variables. Let us define the coefficient
$\lambda$, with the dimensionality  $[length]^{-1}$, according to
\be
p = - \frac{n}{(n^2-1)\,\lambda^2}\,,
\n{lambda}
\ee
and introduce a new dimensionless quantity $\xi=\lambda r$
and a new dimensionless function, $Q(\xi)$, according to
\be
\nu(\xi) = \lambda^{2-n}\, Q(\xi)\,.
\ee
Notice that the dimensionality of $\nu$ is $[length]^{n-2}$,
while $Q$ is dimensionless.
The master equation (\ref{master nu}) can be cast into the
form
\ba 
&& 4 \xi^3 {d^3{Q}\over d\xi^3}  - 4 \xi^2 (n-1)  {d^2 Q\over d\xi^2}
- \xi  \left(\xi ^2+4 n+8\right)  {dQ\over d\xi}
\nonumber
\\
&+& \left(4n + 8 - \xi ^2\right) Q\,=\, 0\,.
\n{master Q}
\ea 
The last equation can be rewritten as
\be
4\xi^2 \frac{d^2T}{d\xi^2}  - 4 \xi(n+2) {dT\over d\xi}  
+ (4n + 8 - \xi^2) T\,=\,0\,,
\n{master T}
\ee
where
$$
Q(\xi) = \frac{S(\xi)}{\xi}\quad \mbox{and}\quad
{dS\over d\xi}=T(\xi)\,.
$$
The equation (\ref{master T}) can be easily solved in
terms of the modified Bessel functions $K_{q}(z)$ and
$I_{q}(z)$. The final solution for  $Q(\xi)$ has the form
\ba
Q(\xi) &=&
\frac{C_0}{\xi}
\n{eqn Q}
\\
&+& \xi^{(n+1)/2} \Big[C_1 K_{\frac{n+3}{2}}({\xi/2})
+ \tilde{C}_2I_{\frac{n+3}{2}}({\xi/2})\Big]\,.
\nonumber
\ea

Since we assume that at infinity the spacetime is flat we
have to put one of the integration constants zero,
$\tilde{C}_2=0$. In this way we arrive at the final
form of the physically interesting solution
\be
Q(\xi)\,=\,\frac{C_0}{\xi}
+ \xi^{(n+1)/2} \Big[C_1 K_{\frac{n+3}{2}}(\xi/2)\Big]\,.
\n{eqn Q int}
\ee
For completeness, we present also the same solution
in terms of original variables
\be
\n{eqn}
\nu\,=\,
\frac{\tilde{C}_0}{r}
+ \tilde{C}_1 r^{(n+1)/2}  K_{\frac{n+3}{2}}
(\lambda r/2)\,.
\ee

Let us now find the expressions for $\alpha$ and $\beta$.
For this end we can use the results of the previously
considered procedure, which involved the equations (\ref{eq1}),
(\ref{eq2}), (\ref{eq3}) and led to the master equation
(\ref{master nu}). Solving equations (\ref{eq1}),
(\ref{eq2}), (\ref{eq3}) we find an algebraic
solution for $\alpha$. After using (\ref{eqn Q int}),
we arrive at the following result:
\ba
\alpha(\xi)&=&
\frac{1}{\xi^n\,n (n+1)}\,\Big[
4 \xi ^2 \frac{d^2 Q}{d\xi^2} - (\xi^2 + 8) Q\Big] 
\n{alpha}
\\
&=&
- \frac{C_0}{n(n+1)}\,\xi^{1-n}
\,-\,{2 C_1\over n+1} \xi^{(3-n)/2}K_{\frac{n+1}{2}}(\xi/2)\, .
\nonumber
\ea

Equation \eq{eq2} leads to the equation for $\beta$,
\ba
{d\beta\over d\xi}&=&
-\, \frac{2}{\xi^n\,(n+1)}\,\Big[
\xi \frac{d^2 Q}{d\xi^2}  + 2 \frac{d Q}{d\xi}\Big] 
\nonumber 
\\
&=&{C_1\over 2(n+1)}\xi^{\frac{1-n}{2}} 
\left[ 2K_{\frac{n+1}{2}}(\xi/2)
-\xi K_{\frac{n-1}{2}}(\xi/2)\right]\, .
\nonumber 
\ea
Integration of this relation gives
\ba
\beta(\xi) &=&C_2
\\
&+& {C_1\over n+1}\xi^{\frac{1-n}{2}}
\left[\xi
K_{\frac{n+1}{2}}(\xi/2) - 2nK_{\frac{n-1}{2}}(\xi/2)\right]\,.
\nonumber 
\ea
Substitution of the obtained solutions for 
$\nu(\xi)$, $\alpha(\xi)$,
and $\beta(\xi)$ into the original set of the equations
\eq{eq1}--\eq{eq3} enables one to make an additional check 
of the correctness of the obtained solutions.

The modified Bessel functions $K_{\mu}(z)$, which enter the 
expressions for $\nu$, $\alpha$ and $\beta$, have the
following asymptotic behavior at the infinity
\be\n{asym}
K_{\mu}(z)\sim 
\sqrt{ {\pi\over 2z}} e^{-z}\left( 1+{4\mu^2 -1\over 8 z}
+ \ldots\right)\, .
\ee
Notice, that at the leading order these asymptotics are 
universal, that is they do not depend on the index $\mu$.
Using \eq{asym} we obtain the following asymptotic 
expressions at large $\xi$ for the functions $\nu$, $\alpha$ 
and $\beta$ 
\ba
\nu &=& \lambda^{2-n} Q \hh
Q \sim {C_0\over \xi} 
+ C_1 \sqrt{\pi} \,\xi^{n/2} \,e^{-\xi/2}\,,
\\
\alpha &\sim& 
- {C_0\over n(n+1)}\, \xi^{1-n} 
- \frac{2\sqrt{\pi}C_1}{n+1}\,\xi^{1-n/2}\, e^{-\xi/2}\,,
\\
\beta &\sim& C_2
+ \frac{\sqrt{\pi}}{n+1}\,\xi^{1-n/2}\, e^{-\xi/2}\,.
\ea

\section{Finding a solution by iterations}

Now we study solutions of the system \eq{vr}--\eq{P}
without making an assumption that the gravitation field
is weak. As we already mentioned in Section~2 the 
corresponding black hole solutions has a characteristic 
scale connected with its gravitational radius $r_0$. 
For example, in the absence of quantum corrections the
corresponding (Tangherlini) metric is
\be
ds^2=-[1-(r_0/r)^{n-1} ]dv^2+2 dv dr+ r^2 d\omega^2_n\, .
\ee
One can always rescale this metric as follows
\be
ds^2=r_0^2 dS^2\, ,
\ee
where $dS^2$ is the dimensionless form of the metric. Making 
the calculations in this dimensionless form is equivalent 
putting $r_0=1$ in the `physical' metric $ds^2$. In what 
follows we always assume $r_0=1$. It is convenient to rewrite 
the main system of equations \eq{vr}-\eq{P} in terms of the 
new coordinate $x=1/r$. We also introduce the 
new functions $M(x)$ and $\Psi(x)$ according to
\be
A=1-x^{n-1} M(x)
\hh
\Psi=\dot{C}\,.
\ee
We denote by a `dot' a derivative with respect to the
coordinate $x$. Our normalization means that the asymptotic 
values of $M$ at $x\to 0$ is always equal to 1.

In these coordinates and notations the system \eq{vr}--\eq{P}
takes the form
\ba
\dot{M} &=& {p(1-n)\over n }\left[-2x^n\{
[(n+1)\dot{K}+2x\ddot{K}]M+x\dot{K}\dot{M}\}\right.\nonumber\\
&+&
\left.
4(x^2\ddot{K}+x\dot{K}-K)+x^{n-2}K^2\right]\,,
\n{M} \\
\Psi&=& -{2p(n-1)\over n}x^{n+1}[\ddot{K}-\dot{K}\Psi]\,,
\n{Ps} \\
K &=& (n+1)[nxM+2x^2\dot{M}+2x^2\Psi M-2x^{3-n}\Psi]
\nonumber\\
&-&{p(n-1)(n-2)\over n}x^nK^2\,.
\n{Ka}
\ea
In order to solve this system for small $p$ we use the
following procedure. We start with a `classical' solution.
Thus we put  $p=0$ and obtain
\be
\n{class}
M_0=1\hh \Psi_0=0\hh K_0=n(n+1)x\, .
\ee
One can replace these expressions into \eq{M} and \eq{Ps} 
to obtain $M_1$ and $\Psi_1$. After this, using \eq{Ka} we 
obtain $K_1$ in which all the terms higher that the first 
order in $p$ are omitted. It is possible to repeat this 
procedure to obtain an approximate solution up to any 
order in powers of $p$, by means of iterations.

Denote by $\{M_k,\Psi_k,K_k\}$ the result of the $k$-th
iteration. Then one has
\ba
M_{k+1} &=& 1+p\int_0^x dx F_{M}(M_k,K_k)\,,
\\
\nonumber
\\
\Psi_{k+1} &=& p F_{\Psi}(\Psi_k,K_k)\,,
\\
\nonumber
\\
K_{k+1} &=& F_{K}^{(0)}(M_{k+1},\Psi_{k+1})
+ pF_{K}^{(1)}(K_{k})\,.
\ea
Here
\ba
&& F_{M}(M_k,K_k)
\nonumber
\\
&=&
{(n-1)\over n }\left[
2x^n\{ (n+1)\dot{K}_kM_k - 2x\ddot{K}_k M_k
- x\dot{K}_k\dot{M}_k \}
\right.
\nonumber
\\
&&
\left.
- 4(x^2\ddot{K}_k+x\dot{K}_k-K_k)- x^{n-2}K^2_k
\right]_{\bf [\leq k]}\,,
\nonumber
\ea 
\ba
F_{\Psi}(\Psi_k,K_k)
&=& - {2(n-1)\over n}x^{n+1}
\big[\ddot{K}_k-\dot{K}_k\Psi_k\big]_{\bf [\leq k]}\,,
\nonumber
\ea 
\ba
&& F_{K}^{(0)}(M_{k},\Psi_{k})
\,=\,
(n+1) \big[nxM_k+2x^2\dot{M}_k
\nonumber
\\
&+&
2x^2\Psi_k M_k-2x^{3-n}\Psi_k \big]_{\bf [\leq k+1]}\,,
\nonumber
\ea 
\ba
F_{K}^{(1)}(K_{k})
&=&
{(n-1)(n-2)\over n}\,\big[x^nK_k^2\big]_{\bf [\leq k]}\, ,
\ea
In the formulas above the subscript symbol
$\,_{\bf [\leq k]}\,$ means the expansion into
power series up to the order $k$ in the parameter $p$.
This prescription provides the correct run of the iteration
procedure, such that the $M_k,\,\Psi_k$ and $K_k$
include powers of $p$ up to the order $k$.

The results of the analytical calculations of the first
two iterations for $M$, $\Psi$ and $K$ obtained by using
the Maple program and checked by using Mathematica
are given below:
\ba
M_2 &=&
1\,-\,(n-2) (n-1) (n+1) p x^{n+1}
\label{M2}
\\
&-& \frac{2}{n}\,(n-2)(n^2-1)^2\,p^2
\,\big[(4n^2 + 15n + 6)x^{2n+2} 
\nonumber
\\
&-& 4(n+1)(2n+1)x^{n+3}\big]\,,
\nonumber
\ea
\ba
\Psi_2 = \frac{4p^2 x^{2n+1}}{n} (n-1)^2(n+1)^3(n^2-4)(2n+1),
\label{Psi2}
\ea
\ba
K_2 &=&  n(n+1)x 
\nonumber
\\
&-&  2(n+1)^2(n-2)(n-1)(2n+1) x^{n+2}\,p
\label{K2}
\\
&-& \frac{2}{n}\,(n-2)(n-1)^2(n+1)^3 \,p^2\times
\nonumber
\\
&\times& \big[ \left(8 n^3+69 n^2+66 n+16\right)x^{2n+3}\,
\nonumber
\\
&-& \,8(n+1)(n+2)(2n+1) x^{n+4}\big]\,.
\ea
It is easy to see that the solution presented above
include zero order terms (which are of course the same
as in \eq{class}), first order and second order terms,
all of them have corresponding powers of the expansion
parameter $p$.
It is not difficult to perform further iterations, but
starting from the third order the formulas become very
bulky and hence we do not show them. (The third order
iteration results for $M_3$, $\Psi_3$, and $K_3$ can
be found in the Appendix.)
\vskip 5mm

\begin{figure}[tp]
\begin{center}
\includegraphics[height=5cm]{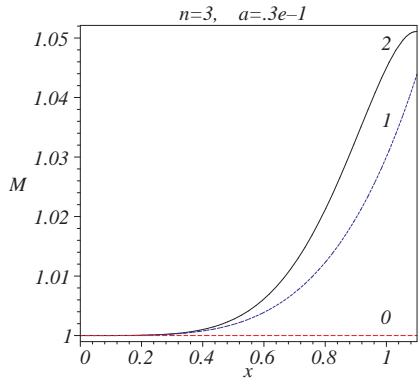}
\caption{Iterations of the mass function $M$ up to the second order. 
The lines $0$, $1$, and $2$ show the functions $M_0(x)$, $M_1(x)$ 
and $M_2(x)$, respectively. The plot is constructed for $n=3$ and 
$a=0.03$.}
\label{fM}
\end{center}
\end{figure}
\begin{figure}[tp]
\begin{center}
\includegraphics[height=5cm]{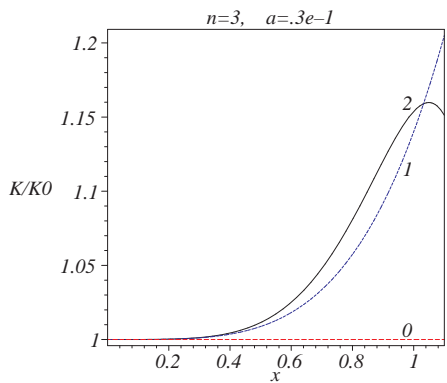}
\caption{Iterations of $K/K_0$ up to the second order. Here 
$K_0=n(n+1)x$ is the value of $K$ for $p=0$. The line $0$ is 
zero order in $p$ result. Lines  $1$ and $2$ show the functions
$K_1(x)/K_0(x)$ and $K_2(x)/K_0(x)$, respectively. The plot 
is constructed for $n=3$ and $a=0.03$.}
\label{fK}
\end{center}
\end{figure}
\begin{figure}[tp]
\begin{center}
\includegraphics[height=5cm]{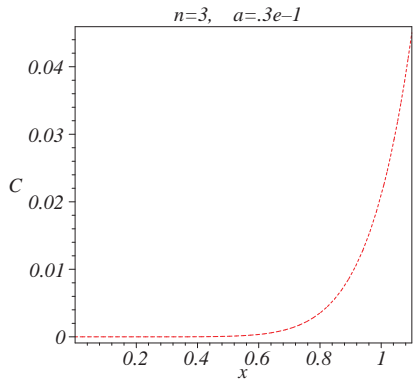}
\caption{Iterations of $C(x)$ up to the second order. The functions
$C_0$ and $C_1$ vanish, so that the plot shows only $C_2$ function.}
\label{fC}
\end{center}
\end{figure}

The plot in Figure \ref{fM} shows that the mass function $M(x)$ 
grows from its value $1$ at infinity to a larger value at the 
horizon. A natural interpretation of this result is that 
introducing the the Weyl term to the action is qualitatively 
equivalent to the addition of a negative mass density 
distribution into the black hole exterior. Let us remember 
that we assume that the Weyl term in the action produces 
the change of the black hole solution which is close to 
the one of the vacuum polarization effect.

We use equation
\be
M(x_H)=x_H^{1-n}\, ,
\ee
to obtain a position of the horizon. We also use the 
expression (\ref{kappa}) for the surface gravity
\be
\kappa \,=\, -\frac12\, [x^2 e^C \dot{A}]_{x=x_H}\, .
\ee
The results of the calculations of $x_H$ and $\kappa$ are given 
in the Table.
\vskip 2mm

\begin{table}[h]
\begin{center}
\begin{tabular}{|c|c|c|c|c|}        \hline
$a$ & & $0.01$ & $0.02$ & $0.03$ \\ \hline
 $x_H$ & $n=3$	& $ 0.994 $  & $ 0.987 $& $ 0.979 $ \\ \hline
$x_H$ & $n=4$	& $ 0.987 $  & $ 0.970 $& $ 0.950 $ \\ \hline
 $\kappa/\kappa_0$ & $n=3$	& $ 1.014 $& $ 1.028 $& $ 1.041 $ \\ \hline
 $\kappa/\kappa_0$ & $n=4$	& $ 1.036 $& $ 1.074 $& $ 1.124 $ \\ \hline
\end{tabular}
\end{center}
\label{tab}
\caption[t1]{\sl 
Position of the horizon, $x_H$, and the surface gravity
$\kappa$ of distorted black holes.}
\end{table}
\vskip 4mm

This table shows that, as expected, for smaller values of $a$,
the solution are closer to their unperturbed ones. The size of the
event horizon, $r_0/x_H$, is always greater than its unperturbed
value $r_0$.
As far as the value of $a$ increase, the gravitational radius
$r_H=x_H^{-1}$ grows up and the surface gravity grows up as well.

\section{4D case}

As we already mentioned in the Introduction, the $4D$-case is a
special one. Let us consider it in more details, using the results
of the iteration procedure presented in the previous section.

The iteration formulas 
\eq{M2}, \eq{Psi2} and \eq{K2} imply that in the
4-dimensional case, when $n=D-2=2$, the corrections to the
`classical' solution \eq{class} vanish. One can check this
directly by substituting \eq{class} into the equations
\eq{M}-(\ref{Ka}). In principle, there might exist
non-perturbative in $p$ solutions close to the Schwarzschild
metric. Let us demonstrate that at least for small
perturbations which can be treated in the linearized
approach, such regular both at the infinity and at the horizon
solutions do not exist. For this purpose let us consider small 
perturbations near the `classical' solution
\be
M=1+\epsilon m(x)
\hhh
\Psi= \epsilon v(x)
\hhh
K=6x\big[ 1 + \epsilon k(x) \big]\,.
\ee
Here $\epsilon$ is a small parameter which should be set 
to unity after performing the series expansion.
Substituting these relations into the basic equations
\eq{M}-(\ref{Ka}) one obtains that the  terms of
the order of $\epsilon^0$ vanish. Keeping
leading linear in $\epsilon$ terms one obtains a following set of
equations
\ba
&&6px^2 [2 x(1-x)\ddot{k}+(6-7x)\dot{k}+3k]
\nonumber
\\
&&+(1-6px^3)\dot{m}-18px^2 m=0\, ,\n{p1}
\\
\nonumber
\\
&&v+6px^3[x\ddot{k}+2\dot{k}-v]=0\, ,\n{p2}
\\
\nonumber
\\
&&m+x\dot{m}-k-(1-x)v=0\, .\n{p3}
\ea

We obtain now a master equation for this system. For this purpose we
use \eq{p2} to find $v(x)$
\be
v \,=\,\frac{6p x^3}{6px^3-1}\,\big(x\ddot{k} + 2\dot{k}\big)\,.
\n{pv}
\ee
Let us substitute this expression into \eq{p3}. Using the obtained
equation and \eq{p1} one can solve these linear with respect to $m$
and $\dot{m}$ equations to obtain the both quantities. One gets
\ba
m &=&
 k + {6px^3\over 1+12px^3}\left[
x(1-x)\ddot{k}+(4-5x)\dot{k}\right]\, ,\n{m1}
\\
\dot{m} &=&
\frac{6px^2}{(1 - 6px^3)(1+12 p x^3)}
\Big[ 2x (x-1)(3px^3 + 1)\ddot{k}
\nonumber
\\
&+& (7x - 6px^4 - 6)\dot{k}
\Big]
\,.\n{m2}
\ea

Differentiating \eq{m1} and putting it equal to $\dot{m}$ defined by
\eq{m2} one obtains a master equation for $k$. From the structure of
the expressions for $m$ and $\dot{m}$ it is evident that the master
equation does not contain $k$, but only its derivatives.
This means that $k=const$ is a solution. This solution results in a
simple change of the gravitational radius $r_0$ which we originally
put equal to 1. We do not consider this renormalization ambiguity and
in what follows will keep $r_0=1$. 

Let us now study possible non-trivial perturbed solutions.
By denoting $\dot{k}=Y$ we write the master equation in the form
\be
F_2 \ddot{Y}+F_1\dot{Y}+F_0 Y=0\, ,
\n{FY}
\ee
\ba
F_0&=&1+108px^2-144p x^3-108 p^2 x^6+1728 p^3 x^9\, ,\nonumber\\
F_1&=&12 p x^3 (5-6x+21 p x^3-27 p x^4 \nonumber\\
&&\hspace{1cm}-144 p^2 x^6+216 p^2 x^7)\, ,\nonumber\\
F_2&=&6 p x^4 (1-x)(1+12px^3)(1-6px^3)\, .
\ea

As any second order ODE, \eq{FY} can be
written in a self-adjoint form
\be
{d\over dx}\left(f_1 {dY\over dx}\right)+f_0 Y=0\, .\n{fY}
\ee
By comparing \eq{FY} and \eq{fY} we have
\be
{\dot{f}_1\over f_1}={F_1\over F_2}\hh
{{f}_0\over f_1}={F_0\over F_2}\, .\n{fF}
\ee
Integrating the first of these equations we obtain
\be
f_1={ (1-x)^2 x^{10}\over (1-6px^3)(1+12px^3)}\, .
\ee
A general solution contains as a common factor an arbitrary
integration constant $C$. We put this constant  equal to
1. With this choice and $p>-1/12$, $f_1(x)$ is positive in the 
black hole exterior. Using the second equation of
\eq{fF} we get
\ba
f_0 = { (1-x) x^6 Q\over
6p(1-6px^3)^2(1+12px^3)^2}\, ,
\ea
\ba
Q = 1+108 p x^2-144 p x^3-108 p^2 x^6+1728 p^3 x^9\, .
\ea

The horizon, $x=1$, and the infinity, $x=0$, are singular points of
the equation \eq{fY}. Let us consider first this equation in the
vicinity of the horizon. Denote $x=1-y$, then in the region $y\approx
0$ one has
\be
y{d^2Y\over dy^2}+2{dY\over dy}+cY=0\hh
c={1-30p-288p^2\over 6p(1+12p)}\, .
\ee
Putting $Y\sim y^{\alpha}$ one finds that $\alpha(1+\alpha)=0$. 
It means that one of the two linearly independent solutions is 
singular, $Y\sim y^{-1}$,  at this point. Hence only one of 
these two solutions is regular at this point, and it is uniquely 
determined by the boundary condition $Y(y=0)=Y_0$.

Let us analyze solutions of \eq{fY} near the infinity, that is 
near $x=0$. In the close vicinity of this point one has
\be
F_2\sim 6p x^4\hh
F_1\sim 60 p x^3\hh
F_0\sim 1\, ,
\ee
and the equation \eq{FY} takes the form
\be
6p x^4 \ddot{Y}+60px^3 \dot{Y} +Y=0\, .
\ee
Changing the
coordinates $r=1/x$ and keeping leading at $r\to \infty$ terms we
obtain the following asymptotic form of the equation
\be\n{RINF}
6p {d^2Y\over dr^2}-48p {dY\over dr}+r Y=0\, .
\ee
The asymptotic form of the solutions at infinity is
\be
Y\sim \exp (\pm
r/2\lambda )\hh \lambda= \sqrt{-6p}\, .
\ee
A solution of \eq{RINF} which decreases at infinity is
\be
Y=C (r^4 +10 \lambda r^3+45 \lambda^2 r^2+105 \lambda^3 r+105
\lambda^4)e^{-r/\lambda }\, .
\ee
Since a regular solution is uniquely fixed by its value at the
horizon, in a general case such a solution would become increasing at
infinity. Thus in a general case $Y=0$ is the only solution which is
regular both, at the horizon and infinity. Based on this analysis, one
however cannot exclude the existence of regular solutions for some
discrete  values of $p$. We demonstrate now that it never happens when
$p>-1/12$.

\begin{figure}[tp]
\begin{center}
\includegraphics[height=5cm]{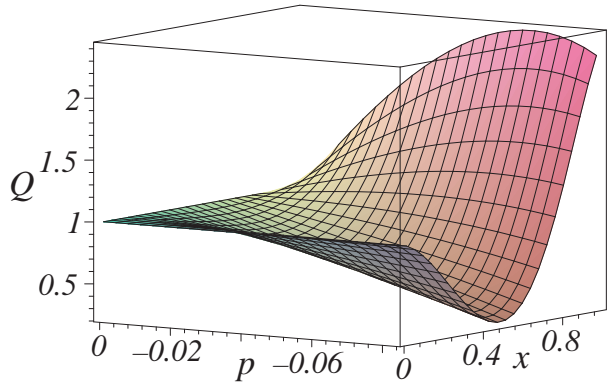}
\caption{Function $Q(x,p)$ in the domain $x\in[0,1]$, $p\in
[0,-1/12]$. It is clear from this plot that this function is positive
everywhere in this domain.}
\label{fQ}
\end{center}
\end{figure}

For this purpose
let us multiply \eq{fY} by $Y$ and integrate the obtained expression
in the interval $x\in (0,1)$. After integration by parts one has
\be
\int_0^1 dx [f_1\dot{Y}^2-f_0 Y^2]=[f_1\dot{Y} Y]^1_0\, .
\ee
For a solution which decreases at infinity and is bounded at the
horizon $x=1$, where $f_1(1)=0$, the expression in the right-hand side vanishes.
For $p<-1/12$ the function $f_1(x)$ is positive in the interval
$x\in(0,1)$.
Numerical analysis shows (see Figure~\ref{fQ}) that
for $p<-1/12$   the function
$f_0(x)$ is negative in the interval $(0,1)$. Hence the integrand in
the left hand side is non-negative everywhere in the interval $x\in
(0,1)$. Hence both $Y$ and $\dot{Y}$ must vanish identically. This
proves that for $p<-1/12$ the equation has only one bounded solution,
$Y(x)=0$. It means that for small $p$ the Schwarzschild metric is the
only one solution of the modified by ${\cal C}^2$ correction Einstein
gravitational equations (at least in the close vicinity of this
solution).

\section{Discussions}

We study black hole solutions in the theory with the Weyl action
correction to the Einstein gravity in higher dimensional spacetimes.
We demonstrated that in the higher dimensional case with $D\ge 5$ 
this correction results in the modification of the Tangherlini 
metric, which is uniquely determined the one parameter, the 
value of the gravitational radius. We developed the iteration 
procedure which for a small value of $\alpha=a/r_0^2$ allows 
one to obtain a solution outside the horizon. This iteration 
procedure does not work uniformly. In particular for a given 
value of $\alpha$ there always exist such a value of $r$ where 
the iteration is not convergent. For small enough value of 
$\alpha$ this value of $r$ is inside the gravitational radius. 
In this domain terms with higher in curvature corrections must
play an important role. It is an interesting problem to develop a
method of solving the equations in this domain either by obtaining a
reliable analytic approximation or by developing stable numerical
schemes. Another interesting open question is to study uniqueness of
the exterior solution obtained by the iteration method. We analyze a
similar problem in the $4D$-case where the situation is much simpler.
We first demonstrate that the classical Schwarzschild metric is a
fixed point of the iteration procedure. After this we studied
metrics in the vicinity of the Schwarzschild metric and demonstrated
that at least for the value of $\alpha <1/2$ ($p>-1/12$) such regular
perturbative solutions are absent. We also showed that in a general
case  additional to the Schwarzschild
solutions do not exist. It is interesting to generalize these results
to the higher dimensional case. It should be emphasized, that the 
action with the ${\cal C}^2$ term does not modify the $4D$ black 
hole solution. However a similar to the higher
dimensional case happens, for example, if one considers a correction
to the Einstein action of the form $f({\cal C}^2)$, or even more
general functions of the curvature invariants.

\vspace{1.5cm}
\section*{Acknowledgments}
V.F. thanks the Natural Sciences and Engineering Research 
Council of Canada and the Killam Trust for the support. 
The work of I.Sh. was suported by CNPq, FAPEMIG, FAPES and 
ICTP. The work on this paper started during the visit of 
V.F. to Brazil. Authors are grateful to FAPEMIG (Minas 
Gerais, Brazil) for partial support of this visit. Also,
V.F. is grateful to the Physics Department of the Federal 
University of Juiz de Fora for hospitality.

\section*{Appendix. Results of iterations in the third order}

Here we present the third order results for the iteration
procedure developed in the section 6.
\begin{widetext}
$$
M_3 \,=\, 1
\,-\,(n-2) (n-1) (n+1) p x^{n+1} 
-\frac{2}{n}\,(n-2)(n^2-1)^2 \,p^2\,
\Big[(4n^2 + 15n + 6)x^{2n+2} -  4(n+1)(2n+1)x^{n+3} \Big]
\qquad\qquad\qquad\qquad\qquad\qquad\qquad\qquad\quad
$$$$
-\frac{ 4(n-2)(n^2-1)^3}{3 n^2}\, p^3\,
 \Big[
(68 n^4 + 745n^3 + 1699n^2 + 1166n + 240) x^{3n+3}
\qquad\qquad\qquad\qquad\qquad\qquad\qquad\qquad
$$$$
- 12(n+1)^2(12n^2 + 97n + 44) x^{2n+4}
+ 48(n+1)(n+2)(n+3)(2n+1) x^{n+5}
\Big] \,.\qquad\qquad\qquad\qquad\qquad
\eqno(A1)
$$
\vskip 2mm

$$
\Psi_3\,=\,\frac{4(n-1)^2(n+1)^3(n^2-4)(2n+1)}{n}\,p^2\,
\, x^{2 n+1} 
\qquad\qquad\qquad\qquad\qquad\qquad
\qquad\qquad\qquad\qquad\qquad\qquad\qquad\qquad\quad
$$$$
+ \frac{32 (n-2) (n-1)^3 (n+1)^4}{n^2}\, p^3\,
\Big[
  (4n^4 + 41n^3 + 86n^2 + 58n + 12)x^{3n+2}
- (n+2)(n+3)(n+4)(2n+1) x^{2n+3}
\Big]\,.
\eqno(A2)
$$
\vskip 2mm

$$
K_3\,=\, n(n+1)x
\,-\,2 (n+1)^2(n-2)(n-1)(2n+1) \, p \,x^{n+2}
\qquad\qquad\qquad\qquad\qquad \qquad\qquad\qquad\qquad\qquad
\qquad\qquad\quad
$$$$
-\,\frac{2 (n-2) (n-1)^2 (n+1)^3}{n}\, p^2 \,
\Big[
\left(8 n^3+69 n^2+66 n+16\right)x^{2n+3}\,
-\,8(n+1)(n+2)(2n+1) x^{n+4}\Big]
$$$$
- \,\frac{8(n-2)(n-1)^3(n+1)^4}{3 n^2}\, p^3\, \,
\Big[ 48 (n+3)^2(n+1)(n+2)(2n+1) x^{n+6}
\qquad\qquad\qquad\qquad
$$$$
- 6(n+1)(32n^4 + 399n^3 + 1005n^2 + 828n + 208) x^{2n+5}
\qquad\qquad\qquad\qquad\qquad\qquad
$$$$
+ (142n^5 + 1643n^4 + 5222n^3 + 5878n^2 + 2700n + 432) x^{3n+4}
\Big]\,\,.
\qquad\qquad\qquad\qquad
\eqno(A3)
$$
\end{widetext}



\begin{thebibliography}{9}

\bibitem{Pope} 
W. Chen, H. L\"u and C. Pope,  {\em Class. Quantum Grav.}
{\bf 23}, 5323 (2006).

\bibitem{Tang} F.R. Tangherlini,  \NC 27, 636 (1963).

\bibitem{DW} DeWitt B. S. {\em Dynamical Theory of Groups and Fields},
(Gordon and Breach, New York, 1965).

\bibitem{PoImpo} I.L. Shapiro,
Class. Quant. Grav. 25 (2008) 103001; arXiv: 0801.0216 [gr-qc].

\bibitem{star}
A.A. Starobinsky, Phys.Lett. {\bf 91B} (1980) 99;
{\sl Nonsingular Model of the Universe with the
Quantum-Gravitational De Sitter Stage and its
Observational Consequences,} Proceedings of the
second seminar "Quantum Gravity", pp. 58-72 (Moscow, 1982);
JETP Lett. {\bf 30} (1979) 719;  {\bf 34} (1981) 460;
Let.Astr.Journ. (in Russian), {\bf 9} (1983) 579.

\bibitem{Pol} 
J. Polchinski, {\em String Theory} (Cambridge University
Press, 1998).

\bibitem{Pen} 
R. Penrose, {\em Phys. Rev. Lett.} {\bf 14}, 57 (1965).

\bibitem{Hawk}
S. Hawking, {\em Proc. R. Soc. London} {\bf A 300}, 182 (1967).

\bibitem{PH}
S. Hawking and R. Penrose,{\em Proc. R. Soc. London} {\bf
 A 314}, 529 (1970).

\bibitem{FMM} 
V. Frolov, M. Markov and V. Mukhanov, {\em Phys. Rev.}
{bf D 41}, 383 (1990).

\bibitem{BF} 
C. Barrabes and V. Frolov, {\em Phys. Rev.} {\bf D53}, 3215 (1996).

\bibitem{FN} 
V. P. Frolov and I. D. Novikov, {\em Black Hole Physics: Basic
Concepts and New Developments}, Kluwer Academic Publishers (1998).

\bibitem{birdav} N.D. Birell and P.C.W. Davies,
{\it Quantum Fields in Curved Space}
(Cambridge University Press, Cambridge, 1982).

\bibitem{book}
I.L. Buchbinder, S.D. Odintsov and I.L. Shapiro,
{\it Effective Action in Quantum Gravity}
(IOP Publishing, Bristol, 1992).

\bibitem{FV} 
V.P. Frolov and G.A. Vilkovisky, {\em Physics Letters}
{\bf B 106}, 307 (1981); ICTP preprint IC/79/69 (1979) Trieste.

\bibitem{HigherBH}
R.C. Myers, J.Z. Simon, Gen. Rel. Grav. 21, 761 (1989);

R.C. Myers,
{\em Black holes in higher curvature gravity.}, in
B.R. Iyer (Ed.) et al.: Black holes, gravitational radiation and the
universe, 121,  gr-qc/9811042;

T. Jacobson and R.C. Myers,
Phys. Rev. Lett. 70, 3684 (1993), hep-th/9305016;

T. Jacobson, G. Kang and R.C. Myers,
Phys. Rev. D52, 3518 (1995), gr-qc/9503020;

S. Deser, A.V. Ryzhov, Class. Quant. Grav. 22, 3315 (2005), 
gr-qc/0505039;

A. Sen, JHEP 0509, 038  (2005) hep-th/0506177;

J. Matyjasek, M. Telecka and D. Tryniecki,
Phys. Rev. D73, 124016 (2006); 

W. Berej, J.Matyjasek, D. Tryniecki, M. Woronowicz
Gen. Rel. Grav. 38, 885 (2006); 

R. Aros, M. Contreras, R. Olea, R. Troncoso, J. Zanelli, 
Phys. Rev. Lett. 84, 1647 (2000), gr-qc/9909015;

G. Kofinas, R. Olea, Phys.Rev. D74, 084035 (2006), hep-th/0606253;

G. Kofinas, R. Olea, JHEP 0711, 069(2007), ArXiv: 0708.0782;

C.-M. Chen, D.V. Gal'tsov and D.G. Orlov,
Phys. Rev. D78, 104013 (2008),  arXiv:0809.1720;

C. Garraffo, G. Giribet, Mod. Phys. Lett. A23, 1801 (2008),
arXiv:0805.3575;

R. Biswas and S. Chakraborty, {\em
The geometry of the higher dimensional black hole
thermodynamics in Einstein-Gauss-Bonnet theory},
 arXiv:0905.1776 (gr-qc);

R. Emparan, H.S. Reall, Living Rev. Rel. 11, 6 (2008),
arXiv:0801.3471 (hep-th).

Z.~K.~Guo, N.~Ohta and T.~Torii,
Prog.\ Theor.\ Phys.\  {\bf 120} (2008) 581; 
{\bf 121} (2009) 253; 
959.



\bibitem{CPW1} S. Coleman, J. Preskill, and F. Wilczek, 
Phys. Rev. Lett. {\bf 67}, 1975 (1991).

\bibitem{CPW2} S. Coleman, J. Preskill, and F. Wilczek, 
Nucl. Phys. {\bf B378}, 175 (1992).

\bibitem{MTW} Misner, C.W.,Thorne, K.S., and Wheeler, J.A. {\em
Gravitation} (W.H.~Freeman, San Francisco, 1973).

\bibitem{Stelle}
K.S. Stelle, Phys. Rev. D16, 953 (1977);
Gen. Rel. Grav. 9, 353  (1978).

\bibitem{Stud}
D.F. Carneiro, E.A. Freitas, B. Gon\c{c}alves,
A.G. de Lima and I.L. Shapiro,
Grav. and Cosm. {\bf 40}, 305 (2004); gr-qc/0412113.


\end{thebibliography}
\end{document}